\documentclass[
aps,prd,
12pt,%10pt
nopreprintnumbers,
showpacs,
eqsecnum,
nofootinbib
]{revtex4-1}

\usepackage{graphicx}
\usepackage{amssymb}

\def\Tr{{\rm Tr\,}}

\begin{document}

\title{Hosotani mechanism in the ``color''-singlet plasma}
\author{Nahomi Kan}\email[]{kan@gifu-nct.ac.jp}
\affiliation{National Institute of Technology, Gifu College,
Motosu-shi, Gifu 501-0495, Japan}
\author{Koichiro Kobayashi}\email[]{kobayasi@oshima-k.ac.jp}
\affiliation{National Institute of Technology, Ohshima College,
Suooshima-cho, Yamaguchi 742-2193, Japan}
%\author{Masashi Kuniyasu}\email[]{mkuni13@yamaguchi-u.ac.jp}
\author{Kiyoshi Shiraishi}\email[]{shiraish@yamaguchi-u.ac.jp}
\affiliation{
%Graduate School of Sciences and Technology for Innovation, 
Faculty of Science,
Yamaguchi University, Yamaguchi-shi, Yamaguchi 753--8512, Japan}
\date{\today}
%\date{}

\begin{abstract}
By projecting the partition function on the ``color''-singlet state,
we investigate the Hosotani mechanism in the fermion-gauge boson plasma. The
present toy-model analysis of the one-loop effective potential at finite
temperature shows that the critical temperature of gauge symmetry breaking
increases at higher temperature in the smaller volume.
\end{abstract}

\maketitle

%%%%%%%%%%%%%%%%%%%%%%%%%%%%%%%%%%%%%%%%%%%%%%%%%%%%%%%%%%%%%%%%%%%%%%%%%%%
%%%%%%%%%%%%%%%%%%%%%%%%%%%%%%%%%%%%%%%%%%%%%%%%%%%%%%%%%%%%%%%%%%%%%%%%%%%
%%%%%%%%%%%%%%%%%%%%%%%%%%%%%%%%%%%%%%%%%%%%%%%%%%%%%%%%%%%%%%%%%%%%%%%%%%%
\section{Introduction}
\label{sec1}
%%%%%%%%%%%%%%%%%%%%%%%%%%%%%%%%%%%%%%%%%%%%%%%%%%%%%%%%%%%%%%%%%%%%%%%%%%%
%%%%%%%%%%%%%%%%%%%%%%%%%%%%%%%%%%%%%%%%%%%%%%%%%%%%%%%%%%%%%%%%%%%%%%%%%%%
%%%%%%%%%%%%%%%%%%%%%%%%%%%%%%%%%%%%%%%%%%%%%%%%%%%%%%%%%%%%%%%%%%%%%%%%%%%
Phase transition phenomena in finite temperature systems are found in various
aspects of physics research.
In the unified theory of particle physics, the analysis of finite-temperature field
theory suggests \cite{DJ} that the broken symmetry by the
Brout--Englert--Higgs (BEH) mechanism was restored in the hot early universe.
In other words, it can be said that the division of the present fundamental
interactions  is the result of the phase transition in the early universe.

The theory of the fundamental scalar field, which is indispensable for the
BEH mechanism, is unnatural in the sense of suffering huge quantum corrections.
Accordingly, the gauge-Higgs unified model has been proposed as a candidate giving
the solutions to the naturalness \cite{gh1,gh2,gh3,gh4,gh5,gh6,gh7}.
In the model, the extra-dimensional component of the gauge field plays the role
of the Higgs field. Thus, there is no potential term that corresponds to the
Higgs potential at the tree level in the theory; this is due to the gauge symmetry,
which guarantees naturalness of the theory.
Therefore, in the gauge-Higgs unified model, it is expected that the quantum
effect at one (or more) loop order will bring about a non-trivial vacuum gauge
field.
This mechanism is dubbed as the Hosotani mechanism (or Wilson loop mechanism)
\cite{Hosotani}. It is known that fermionic matter fields with certain suitable
representations 
%and with nontrivial boundary conditions 
are needed for such
symmetry breaking at zero temperature.
The characteristics of the Hosotani mechanism at finite temperature have already
been studied \cite{Shiraishi,HH1,MT}.
It has been pointed out that phase transitions at high temperature always
occur \cite{HH1} (and the transition is first order for matter fields with the
periodic boundary condition in the extra dimension) in the Hosotani
mechanism.

There is another type of phase transition in non-Abelian gauge theory. It is
the quark-hadron phase transition, which is known as hadronization from quark-gluon
plasma (QGP) \cite{Muller}. Although the similar confinement is not assumed in the
gauge-Higgs scenario, non-perturbative effects are expected in the early phase
of the hot universe.%
\footnote{In a different context, the idea of vacuum selection at finite
temperature in supersymmetric grand unified theories, which is attributed to
the difference in degree of freedom of particles, is discussed by the authors of
the papers on ``supercosmology''
\cite{NT,ELR,NOT,CEHNO}.
}
It is also well known that the AdS/CFT method illustrates the reduction of 
the degrees of freedom in QGP in the non-perturbative region \cite{BCM}
in a certain non-Abelian gauge theory.

Being motivated by the reduction in the number of degrees of freedom in
QGP, we revisit the general behavior of Hosotani mechanism at finite temperature
through a toy model in this paper. We adopt the color-singlet hypothesis in the
present analysis.%
\footnote{The projection onto the singlet sector is also utilized in recent
studies on large $N$ Yang--Mills theories at finite
temperature \cite{Sundborg,AMMPR}.} This is a hypothesis that can be considered in
the hadron phase transition from quark-gluon plasma, and it is a hypothesis that
QGP whitens globally owing to the nature of the strong interaction. There are some
debates as to whether this holds true in the analysis of actual QGPs, etc.
However, since it is followed by a well-defined mathematical operation that
expresses the reduction of degrees of freedom using the group integration
technique, it is considered to be a powerful analytical method that can
approximate the aimed aspect of the effective behavior at least
``phenomenologically''.

In the following sections of this paper, we use a toy model with $SU(2)$ symmetry
(instead of a large group of unified models) to find a one-loop effective
potential at finite temperature under the hypothesis of the color singlet.

This paper is organized as follows. In Sec.~\ref{sec2}, we review the
color-singlet hypothesis and associated technique in analytical formulations.
In Sec.~\ref{sec3}, we define the effective potential for the toy model of
$SU(2)$ gauge theory in the background space $R^{D-2}\otimes S^1$. Numerical
calculations of the effective potential in five dimensions ($D=5$) are shown in
Sec.~\ref{sec4} and the possible phase transition is studied. The 
discussion is presented in the last section.
In Appendix, an interesting and useful analysis of the effective potential with
approximations is examined.
%%%%%%%%%%%%%%%%%%%%%%%%%%%%%%%%%%%%%%%%%%%%%%%%%%%%%%%%%%%%%%%%%%%%%%%%%%%
%%%%%%%%%%%%%%%%%%%%%%%%%%%%%%%%%%%%%%%%%%%%%%%%%%%%%%%%%%%%%%%%%%%%%%%%%%%
%%%%%%%%%%%%%%%%%%%%%%%%%%%%%%%%%%%%%%%%%%%%%%%%%%%%%%%%%%%%%%%%%%%%%%%%%%%
\section{The projected partition function for the color-singlet states}
\label{sec2}
%%%%%%%%%%%%%%%%%%%%%%%%%%%%%%%%%%%%%%%%%%%%%%%%%%%%%%%%%%%%%%%%%%%%%%%%%%%
%%%%%%%%%%%%%%%%%%%%%%%%%%%%%%%%%%%%%%%%%%%%%%%%%%%%%%%%%%%%%%%%%%%%%%%%%%%
%%%%%%%%%%%%%%%%%%%%%%%%%%%%%%%%%%%%%%%%%%%%%%%%%%%%%%%%%%%%%%%%%%%%%%%%%%%

In this section, we review the argument of the global color symmetry
in QGP under the color-singlet hypothesis.
Using the technique shown in the present section, we will obtain the effective
potential in a toy model for the Hosotani mechanism in the ``color''-neutral
plasma in the next section.

The QCD is known as a theory describing the strong force and causes the confinement
of quarks and gluons.  The transition from hadronic matter to quark-gluon plasma is
considered to be  a transition from local color confinement to global color
confinement at finite temperature \cite{Muller}.
We should consider the restricted
partition function of the color-singlet state to realize the
global color symmetry \cite{Muller,EGR,RT,Turko,EG1,EG2,GLPZ,GMPZ}.
To this end, first we define a generalized partition function which includes
the generators of the Cartan subalgebra $\hat{C}_\alpha$ in the gauge group as
follows:
\begin{equation}
Z_C(\psi_\alpha)=\Tr[e^{-\beta\hat{H}+i\psi_\alpha\hat{C}_\alpha}]\,,
\end{equation}
where, as usual, $\beta$ is the inverse of the temperature $T$ and $\hat{H}$
denotes the Hamiltonian. 

Next, we introduce the characteristic functions.
The function of the parameter $\psi_\alpha$ specified by the
representation $j$ of the group ($SU(3)$ for QCD) is called as the
characteristic function $\chi_j(\psi_\alpha)$. The characteristic functions satisfy
\begin{equation}
\int d\mu(\psi_\alpha) \chi^*_j(\psi_\alpha)\chi_{j'}(\psi_\alpha)=\delta_{jj'}\,,
\end{equation}
where $d\mu(\psi_\alpha)$ is the invariant measure of the group.
For $SU(2)$, i.e., the two-color case, it is known that 
\begin{equation}
d\mu(\psi)=\frac{\sin^2\frac{\psi}{2}}{2\pi} d\psi\quad (-2\pi\le
\psi<2\pi)\,,
\end{equation}
where $\psi$ is the single variable for the center of $SU(2)$.

We assume that the generalized partition function can be expanded by the
characteristic functions and the characteristic function for the singlet is known
to be unity.
Thus, the restricted partition function $Z$ is finally obtained by projection
onto the color singlet as
\begin{equation}
Z=\int d\mu(\psi_\alpha) Z_C(\psi_\alpha)\,.
\end{equation}

Here, we notice that the Jacobi's imaginary transformation \cite{AS}
\begin{equation}
\frac{1}{\sqrt{4\pi t}}\sum_{n=-\infty}^\infty e^{-\beta^2n^2/(4t)}e^{i2\pi ny}=
\frac{1}{\beta}\sum_{n=-\infty}^\infty
e^{-[\frac{2\pi}{\beta}(n+y)]^2t}
\end{equation}
and the formula in Ref.~\cite{DJ}, we find that the equalities
\begin{eqnarray}
& &\sum_n\ln\left[\frac{4\pi^2(n+y)^2}{\beta^2}+\omega^2\right]\nonumber \\
&=&-\int_0^\infty\frac{dt}{t}\sum_n\exp\left[-\left(\frac{4\pi^2(n+y)^2}{\beta^2}+\omega^2\right)t
\right]=
-\frac{\beta}{\sqrt{4\pi}}\int_0^\infty\frac{dt}{t^{3/2}}\sum_ne^{i2\pi ny}\exp
\left[-\omega^2t-\frac{\beta^2n^2}{4t}\right]\nonumber \\
&=&-\frac{\beta\omega\Gamma(-\frac{1}{2})}{\sqrt{4\pi}}
-\frac{2\beta\omega}{\sqrt{4\pi}}\sum_{n=1}^\infty 2e^{i2\pi ny}
\left(\frac{2}{n\beta\omega}\right)^{1/2}
K_{1/2}(n\beta\omega)=\beta\omega
-\sum_{n=1}^\infty 
\frac{2}{n}
e^{-n\beta\omega}e^{i2\pi ny}\nonumber \\
&=&\beta\omega+2\ln(1-e^{-\beta\omega+i2\pi y})
\end{eqnarray}
hold, up to the terms independent of an arbitrary constant $\omega$. Therefore, we
can express $\ln Z_C(\psi_\alpha)$ as
\begin{equation}
\ln Z_C(\psi_\alpha)=-\frac{1}{2}\Tr\sum_n\ln\left[\frac{(2\pi
n+\psi_\alpha\hat{C}_\alpha)^2}{\beta^2}+\omega^2\right]\,,
\end{equation}
for bosonic fields and
\begin{equation}
\ln Z_C(\psi_\alpha)=\frac{1}{2}\Tr\sum_n\ln\left[\frac{[2\pi
(n+\frac{1}{2})+\psi_\alpha\hat{C}_\alpha]^2}{\beta^2}+\omega^2\right]\,,
\end{equation}
for fermionic fields,
where each trace indicates the sum over possible energy eigenvalue, $\omega$
and all degrees of freedom.
Precisely speaking, the expressions for the generalized partition function here
involves the contribution of the vacuum energy.
Incidentally, the expression is convenient for evaluation of the effective
potential for the Hosotani mechanism.

%%%%%%%%%%%%%%%%%%%%%%%%%%%%%%%%%%%%%%%%%%%%%%%%%%%%%%%%%%%%%%%%%%%%%%%%%%%
%%%%%%%%%%%%%%%%%%%%%%%%%%%%%%%%%%%%%%%%%%%%%%%%%%%%%%%%%%%%%%%%%%%%%%%%%%%
%%%%%%%%%%%%%%%%%%%%%%%%%%%%%%%%%%%%%%%%%%%%%%%%%%%%%%%%%%%%%%%%%%%%%%%%%%%
\section{The $SU(2)$ toy model}
\label{sec3}
%%%%%%%%%%%%%%%%%%%%%%%%%%%%%%%%%%%%%%%%%%%%%%%%%%%%%%%%%%%%%%%%%%%%%%%%%%%
%%%%%%%%%%%%%%%%%%%%%%%%%%%%%%%%%%%%%%%%%%%%%%%%%%%%%%%%%%%%%%%%%%%%%%%%%%%
%%%%%%%%%%%%%%%%%%%%%%%%%%%%%%%%%%%%%%%%%%%%%%%%%%%%%%%%%%%%%%%%%%%%%%%%%%%

Here, we consider the Hosotani mechanism at finite temperature in
the $SU(2)$ gauge theory with massless fermions in the adjoint
representation.  We consider $D$-dimensional spacetime and assume that the
topology of space is
$R^{D-2}\otimes S^1$ and the circumference of the compact dimension is set to $L$.
All the fields obey the periodic boundary condition with respect to the compact
dimension.%
\footnote{Although various bizarre boundary conditions are investigated by many
papers including Ref.~\cite{ST}, we take the simplest condition in the present
paper.}

Although the several conditions towards asymptotic freedom {etc.}~may be a
necessary condition for strong non-perturbative effects implicitly assumed, we will
temporally ignore the conditions in the present toy model.

The vacuum expectation value of the extra dimensional component of the $SU(2)$
gauge field
$\mathbf{A}_y$ is now parametrized as
\begin{equation}
gL\langle\mathbf{A}_y\rangle=\frac{\theta}{2}\left(
\begin{array}{cc}
1 & 0 \\
0 & -1
\end{array}
\right)\,,
\end{equation}
where $g$ is the $SU(2)$ gauge coupling.
Note that the trivial vacuum is associated with $\theta=0$, where the gauge
invariant Wilson loop over $S^1$ becomes $\exp igL\langle\mathbf{A}_y\rangle=I$,
where $I$ is the identity matrix. The residual large gauge symmetry tells the
identification $\theta\sim
\theta+4\pi$. Moreover, when $\theta=2\pi$, it gives $\exp
igL\langle\mathbf{A}_y\rangle= -I$; then the $SU(2)$ symmetry is unbroken.
This is due to the $Z_2$ symmetry in $SU(2)$ and consequently we expect that the
partition function is periodic in $\theta$ with a period $2\pi$.

The restricted partition function after projections is written as
\cite{Muller,EGR,RT,Turko,EG1,EG2,GLPZ,GMPZ}
\begin{equation}
Z(\theta)={%\textstyle 
\int_{-2\pi}^{2\pi}\frac{d\psi}{2\pi}
\sin^2\frac{\psi}{2}}\,\,{%\textstyle
Z_G(\theta,\psi)}\left[Z_a(\theta,\psi)\right]^{N_a}\,,
\end{equation}
where $N_a$ is the number of the adjoint fermion
fields. In this expression, the logarithm of the generalized partition function
$Z_G$ for the gauge bosons and the ghosts is formally given by
\begin{equation}
\ln Z_G(\theta,\psi)=-\frac{D-2}{2}V\sum_{A=1}^3
\sum_n\sum_k\int\frac{d^{D-2}\mathbf{p}}{(2\pi)^{D-2}}\ln\left[\frac{(2\pi
n+C_A\psi)^2}{\beta^2}+\mathbf{p}^2+M_A^2(k)\right]\,,
\end{equation}
where 
\begin{equation}
C_1=+1\,,\quad C_2=-1\,,\quad C_3=0\,,\quad
M_A^2=\left(\frac{2\pi k+C_A\theta}{L}\right)^2\,,
\end{equation}
and the generalized partition function for the adjoint fermions
are written by
\begin{eqnarray}
& &\ln Z_a(\theta,\psi)\nonumber \\
& &=\frac{2^{[D/2]}}{4}V\sum_{A=1}^3
\sum_n\sum_k\int\frac{d^{D-2}\mathbf{p}}{(2\pi)^{D-2}}\ln\left[\frac{[2\pi
(n+\frac{1}{2})+C_A\psi]^2}{\beta^2}+\mathbf{p}^2+M_A^2(k)\right]\,,
\end{eqnarray}
where $[D/2]$ is the Gauss' symbol such that $[4/2]=[5/2]=2$.

In these expressions, the constant $V$ denotes the $(D-2)$-dimensional volume of
the hot plasma system. The definition of the volume of the system is a crucial
problem. 
%R
We can consider the small objects, such as 
false vacuum bubbles, or the cores of the exotic stars, or the fermion droplets
in the universe. 
%R

We should recall that the above expressions include vacuum contributions and
should be regularized by getting rid of divergences, which are irrelevant to
physical quantities.
With the aid of Jacobi's imaginary transformation \cite{AS}
\begin{equation}
\sum_k \exp\left[-\left(\frac{2\pi k+\Theta}{L}\right)^2t\right]
=\frac{L}{\sqrt{4\pi
t}}\sum_k\exp\left[-\frac{L^2k^2}{4t}\right]
e^{-ik\Theta}\,,
\end{equation}
we find that the similar manipulation as in Sec.~\ref{sec2} leads to
\begin{equation}
\ln Z_G(\theta,\psi)=\frac{(D-2)\Gamma\left(D/2\right)}{2\pi^{D/2}}\beta LV
{\sum_{n,k}}'
\frac{1+2\cos\left[k\theta+n\psi\right]}{\left[
\beta^2n^2+L^2k^2\right]^{D/2}}\,,
\end{equation}
\begin{equation}
\ln
Z_a(\theta,\psi)=\frac{2^{[D/2]}\Gamma\left(D/2\right)}{2\pi^{D/2}}\beta
LV {\sum_{n,k}}'
\frac{(-1)^{n-1}\left\{1+2\cos\left[k\theta+
n\psi\right]\right\}}{\left[
\beta^2n^2+L^2k^2\right]^{D/2}}\,,
\end{equation}
where 
the primes on sums indicate the omission of $n=k=0$
in the summations. By this omission, the expressions have become finite
and the regularization has been accomplished.

Each partition function can be divided into two parts, say,
\begin{eqnarray}
\ln Z_G(\theta,\psi)&=&\ln Z_{G0}(\theta)+\ln
Z_{GT}(\theta,\psi)\,,\nonumber \\
\ln Z_a(\theta,\psi)&=&\ln Z_{a0}(\theta)+\ln
Z_{aT}(\theta,\psi)\,\,.
\end{eqnarray}
Here, contributions of vacuum fluctuations are given,
with the use of the Riemann's zeta function $\zeta_R(z)$ and the $D$-th
polylogarithm function
$\mbox{Li}_D(z)$, by
\begin{eqnarray}
\ln Z_{G0}(\theta)&=&\frac{(D-2)\Gamma\left(D/2\right)}{\pi^{D/2}L^D}
\beta LV {\sum_{k=1}^\infty}\frac{1+2\cos k\theta}{k^D}\nonumber \\
&=&\frac{(D-2)\Gamma\left(D/2\right)}{\pi^{D/2}L^D}\beta LV
\left[\zeta_R(D)+\mbox{Li}_{D}(e^{ik\theta})+\mbox{Li}_{D}(e^{-ik\theta})\right]\,,
\label{Hov1}
\end{eqnarray}
\begin{eqnarray}
\ln
Z_{a0}(\theta)&=&-\frac{2^{[D/2]}\Gamma\left(D/2\right)}{\pi^{D/2}L^D}\beta
LV {\sum_{k=1}^\infty}\frac{1+2\cos k\theta}{k^D}\nonumber \\
&=&-\frac{2^{[D/2]}\Gamma\left(D/2\right)}{\pi^{D/2}L^D}\beta
LV\left[\zeta_R(D)+\mbox{Li}_{D}(e^{ik\theta})+\mbox{Li}_{D}(e^{-ik\theta})\right]\,,
\label{Hov2}
\end{eqnarray}
while the finite-temperature parts are written by
\begin{eqnarray}
\ln Z_{GT}(\theta,\psi)&=&(D-2)\beta LV
\left[w_b(0,0)+2w_b(\theta,\psi)\right]\nonumber \\
\ln
Z_{aT}(\theta,\psi)&=&2^{[D/2]}\beta LV
\left[w_f(0,0)+2w_f(\theta,\psi)\right]
\,,
\end{eqnarray}
where
\begin{equation}
w_b(\theta,\psi)\equiv
%\frac{\Gamma(D/2)}{2\pi^{D/2}}
%\sum_{k=-\infty}^\infty\left\{\frac{\cos\left[k\theta+n\psi\right]}
%{\left[\beta^2n^2+L^2k^2\right]^{D/2}}+
%\frac{\cos\left[k\theta-n\psi\right]}
%{\left[\beta^2n^2+L^2k^2\right]^{D/2}}\right\}\nonumber \\
\frac{\Gamma(D/2)}{\pi^{D/2}}\sum_{n=1}^\infty\sum_{k=-\infty}^\infty\frac{\cos
k\theta
\cos n\psi} {\left[\beta^2n^2+L^2k^2\right]^{D/2}}\,,
\label{wb}
\end{equation}
and
\begin{equation}
w_f(\theta,\psi)\equiv
\frac{\Gamma(D/2)}{\pi^{D/2}}\sum_{n=1}^\infty
(-1)^{n-1}\sum_{k=-\infty}^\infty\frac{\cos k\theta
\cos n\psi} {\left[\beta^2n^2+L^2k^2\right]^{D/2}}\,.
\label{wf}
\end{equation}
%R
Incidentally, $w_b$ and $w_f$ coincide with the effective potential for the 
$SU(2)$ gauge theory with torus ($T^2$) compactification with appropriate
boundary conditions \cite{HH2}. In our case, however, the parameter $\psi$ is to be
integrated for the ``color''-singlet projection.
%R

The partition function restricted to $SU(2)$ ``color''-singlet states
for the present model is then given by
\begin{equation}
Z(\theta)=Z_{G0}(\theta)[Z_{a0}(\theta)]^{N_a}\nonumber
\times\frac{1}{2\pi}\int_{-2\pi}^{2\pi} d\psi\,\Bigl[\sin^2{\frac{\psi}{2}}\Bigr]\,
Z_{GT}(\theta,\psi)[Z_{aT}(\theta,\psi)]^{N_a}\,.
\label{integ}
\end{equation} 
Now, the effective potential ${\cal V}(\theta)$ is defined by
\begin{equation}
{\cal V}(\theta)\equiv-\frac{1}{\beta VL}\ln Z(\theta)\,.
\end{equation}
One can find that the effective potential at zero temperature becomes
\begin{equation}
{\cal V}_0(\theta)\equiv-\frac{1}{\beta VL}[\ln
Z_{G0}(\theta)+N_a\ln Z_{a0}(\theta)]\,,
\end{equation}
where $\ln Z_{G0}$ and $\ln Z_{a0}$ have been
given by (\ref{Hov1}) and (\ref{Hov2}), which is well-known effective potential in
the Hosotani mechanism.

We should also notice that the limit of the infinite volume gives
\begin{equation}
{\cal V}(\theta)\rightarrow {\cal V}_0(\theta)-\frac{1}{\beta VL}[
\ln Z_{GT}(\theta,0)+N_a\ln Z_{aT}(\theta,0)]\,,
\end{equation}
because for a large volume,
$\psi=0$ is a stationary point in the integrand of (\ref{integ}).

If $\theta=0$ gives a global minimum of of ${\cal V}(\theta)$, the gauge symmetry
is unbroken, while if the global minimum is located at $\theta\ne 0 ~(\mbox{mod~}
2\pi)$, the $SU(2)$ symmetry is reduced to be $U(1)$. 

%%%%%%%%%%%%%%%%%%%%%%%%%%%%%%%%%%%%%%%%%%%%%%%%%%%%%%%%%%%%%%%%%%%%%%%%%%%
%%%%%%%%%%%%%%%%%%%%%%%%%%%%%%%%%%%%%%%%%%%%%%%%%%%%%%%%%%%%%%%%%%%%%%%%%%%
%%%%%%%%%%%%%%%%%%%%%%%%%%%%%%%%%%%%%%%%%%%%%%%%%%%%%%%%%%%%%%%%%%%%%%%%%%%
\section{Numerical calculations for $D=5$}
\label{sec4}
%%%%%%%%%%%%%%%%%%%%%%%%%%%%%%%%%%%%%%%%%%%%%%%%%%%%%%%%%%%%%%%%%%%%%%%%%%%
%%%%%%%%%%%%%%%%%%%%%%%%%%%%%%%%%%%%%%%%%%%%%%%%%%%%%%%%%%%%%%%%%%%%%%%%%%%
%%%%%%%%%%%%%%%%%%%%%%%%%%%%%%%%%%%%%%%%%%%%%%%%%%%%%%%%%%%%%%%%%%%%%%%%%%%

Now, we shall evaluate ${\cal V}(\theta)$ in this model.
To handle the numerical function, we use the following integral expression:
\begin{equation}
w_b(\theta,\psi)=\frac{1}{2\pi^{D/2}}\int_0^\infty dt\, t^{D/2-1}
\vartheta_3\left({\textstyle\frac{\theta}{2\pi},i\frac{L^2t}{\pi}}\right)
\left[\vartheta_3\left({\textstyle\frac{\theta}{2\pi},i\frac{\beta^2t}{\pi}}\right)-1\right]\,,
\end{equation}
and
\begin{equation}
w_f(\theta,\psi)=-\frac{1}{2\pi^{D/2}}\int_0^\infty dt\, t^{D/2-1}
\vartheta_3\left({\textstyle\frac{\theta}{2\pi},i\frac{L^2t}{\pi}}\right)
\left[\vartheta_4\left({\textstyle\frac{\theta}{2\pi},i\frac{\beta^2t}{\pi}}\right)-1\right]\,,
\end{equation}
where $\vartheta_n(v,\tau)$ is Jacobi's theta function \cite{AS}.

\begin{figure}[ht]
\centering
\includegraphics[width=5cm]
{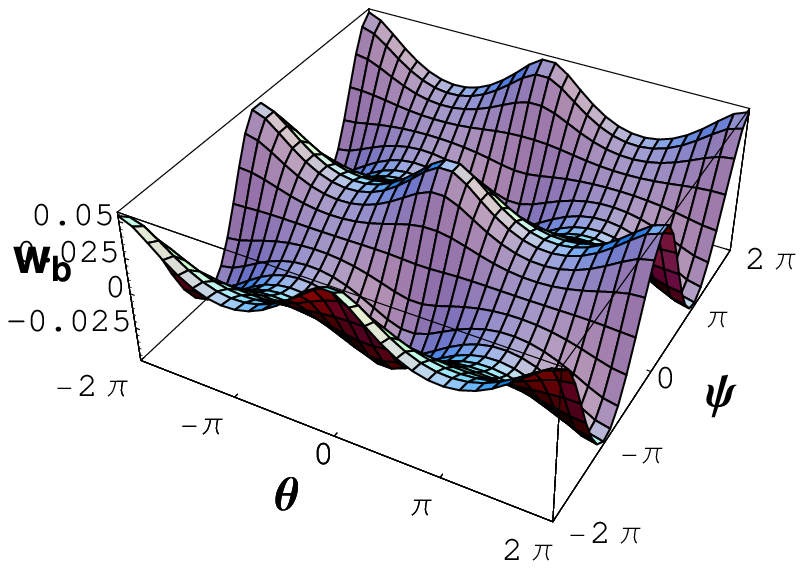}
\hspace{2cm}
\includegraphics[width=5cm]
{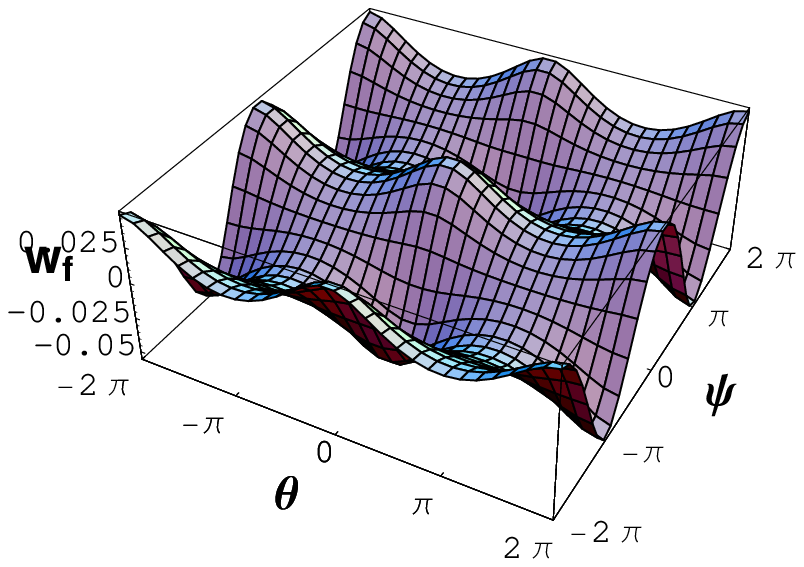}\\
(a) \hspace{7cm} (b)
\caption{The functions for $\beta=1.2$ and $L=1$:  (a) $w_b$ and (b) $w_f$.}
\label{fig1}
\end{figure}

Hereafter, we shall concentrate ourselves on the case with $D=5$.
We show typical shapes of functions $w_b$ and $w_f$ in Fig.~\ref{fig1}.%
\footnote{We used the command \texttt{FunctionInterpolation} in
\textit{Mathematica}
\cite{Mathematica}.}
One can find that both functions $w_b$ and $w_f$ have a period $2\pi$ for both
variables $\theta$ and $\psi$, the maximum of which is at $0$ for each variables
and the minimum of which is at $\pi$. 
Therefore, the extrema of the effective potential ${\cal V}$ should be found at
$\theta=0$ and $\pi$ (modulo $2\pi$).
%%%%%%%%%%%%%%%%%%%%%%%%%%%
% 01
%%%%%%%%%%%%%%%%%%%%%%%%%%%
%\begin{wrapfigure}{r}{5cm}
%\end{wrapfigure}
%%%%%%%%%%%%%%%%%%%%%%%%%%%

\begin{figure}[ht]
\centering
\includegraphics%[width=5cm]
{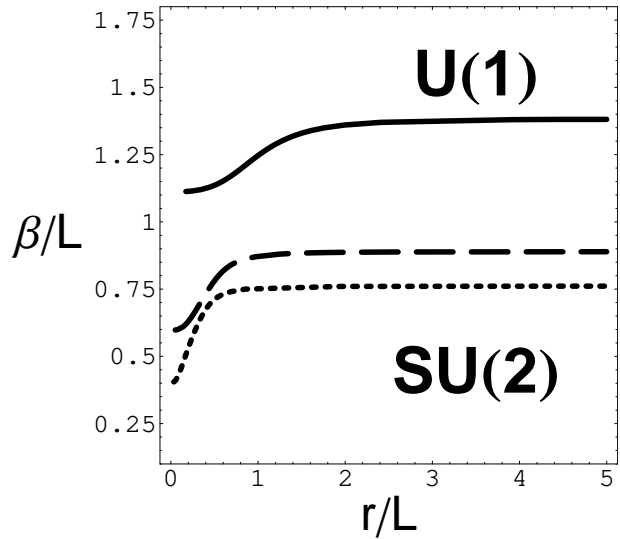}
\caption{Phase structure of the model.
The solid line indicates the boundary of the two phases for $N_a=1$,
the broken line indicates that for $N_a=2$, and
the dotted line indicates that for $N_a=3$.}
\label{fig2}
\end{figure}

At the critical temperature, the values of ${\cal V}(0)$ and ${\cal V}(\pi)$
are equal in the present model.%
\footnote{This is not the case if the matter field with a weird boundary
condition exists in the model.}
In Fig.~\ref{fig2}, we show the critical lines in the
parameter space spanned by $r/L$ and $\beta/L$, 
where, $r$ is the radius of a ball with volume $V$ such that
$V=\frac{4\pi}{3}r^3$.%
\footnote{As is well known,
$V=\frac{\pi^{D/2-1}}{\Gamma\left(\frac{D}{2}\right)}r^{D-2}$
in the case of $D$-dimensional spacetime.}
%%%%%%%%%%%%%%%%%%%%%%%%%%%
% 02
%%%%%%%%%%%%%%%%%%%%%%%%%%%
%\begin{wrapfigure}{r}{5cm}
%\end{wrapfigure}
%%%%%%%%%%%%%%%%%%%%%%%%%%%
In the region above the line, the gauge symmetry
is broken. One can find that the $SU(2)$ symmetry is broken for any value of $r$
for a sufficiently high temperature.
For a smaller $r$, the critical temperature becomes higher; the symmetry
restoration is suppressed by the color-singletness. 
The suppression is larger for a lager number of fermions, which are introduced to
break the symmetry, though it works in a much smaller volume.

\begin{figure}[ht]
\centering
\includegraphics[width=4cm]
{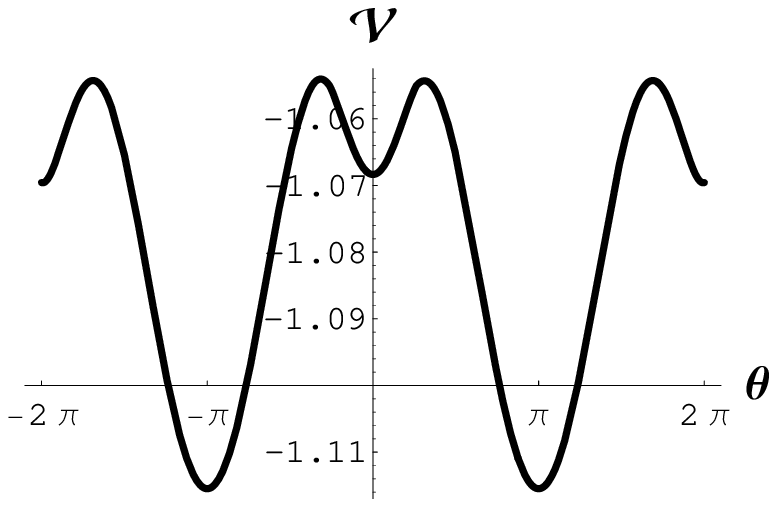}
\includegraphics[width=4cm]
{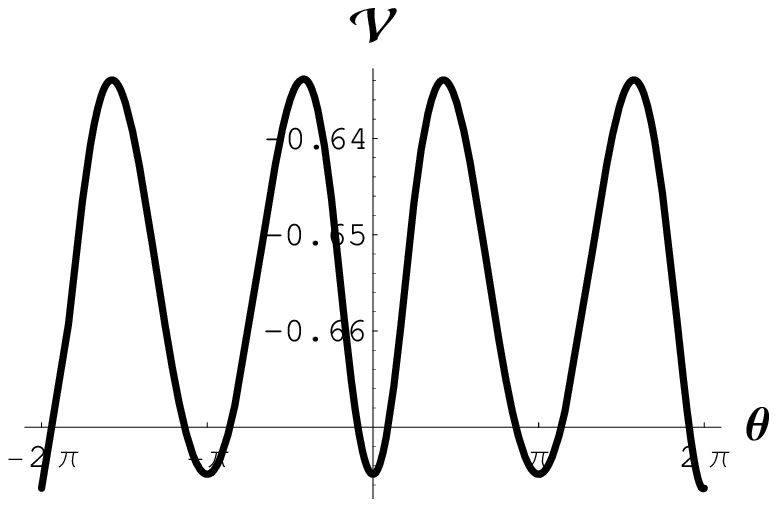}
\includegraphics[width=4cm]
{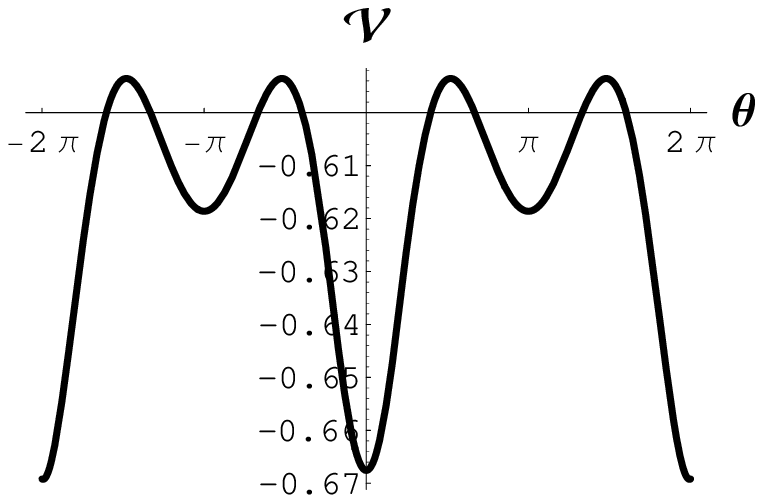}
\includegraphics[width=4cm]
{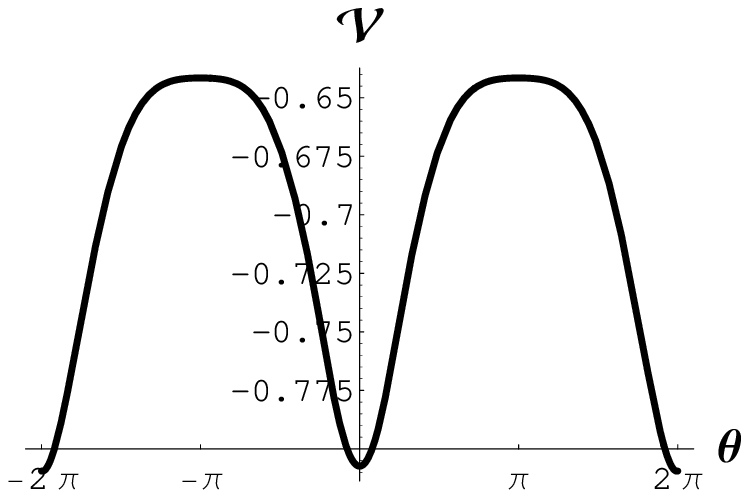}
(a)\hspace{3.5cm}(b)\hspace{3.5cm}(c)\hspace{3.5cm}(d)
\caption{The effective potential ${\cal V}$ for $N_a=1$, $\beta=1.2$ and $L=1$:
(a) $r=0.5$ (b) $r=0.814$ (c) $r=1$ (d) $r=\infty$.}
\label{fig3}
\end{figure}

The typical change in shape of the potential is exhibited in Fig.~\ref{fig3},
which shows the minima of the effective potential in the present model can appear
at $\theta=0$ or $\theta=\pi$ (modulo $2\pi$).
%%%%%%%%%%%%%%%%%%%%%%%%%%%
% 03
%%%%%%%%%%%%%%%%%%%%%%%%%%%
%\begin{wrapfigure}{r}{5cm}
%\end{wrapfigure}
%%%%%%%%%%%%%%%%%%%%%%%%%%%

%%%%%%%%%%%%%%%%%%%%%%%%%%%%%%%%%%%%%%%%%%%%%%%%%%%%%%%%%%%%%%%%%%%%%%%%%%%
%%%%%%%%%%%%%%%%%%%%%%%%%%%%%%%%%%%%%%%%%%%%%%%%%%%%%%%%%%%%%%%%%%%%%%%%%%%
%%%%%%%%%%%%%%%%%%%%%%%%%%%%%%%%%%%%%%%%%%%%%%%%%%%%%%%%%%%%%%%%%%%%%%%%%%%
\section{Discussion}
\label{co}
%%%%%%%%%%%%%%%%%%%%%%%%%%%%%%%%%%%%%%%%%%%%%%%%%%%%%%%%%%%%%%%%%%%%%%%%%%%
%%%%%%%%%%%%%%%%%%%%%%%%%%%%%%%%%%%%%%%%%%%%%%%%%%%%%%%%%%%%%%%%%%%%%%%%%%%
%%%%%%%%%%%%%%%%%%%%%%%%%%%%%%%%%%%%%%%%%%%%%%%%%%%%%%%%%%%%%%%%%%%%%%%%%%%

In this paper, the effective potential at finite temperature was obtained for the
$SU(2)$ toy model based on the color-singlet hypothesis.
For a smaller volume, the critical temperature of the $SU(2)$-$U(1)$ phase
transition becomes higher. 

%%%%%%%%%%%%%%%%%%%%%%%%%%%%%%%%%%%%%%%%%%%%%%%%%%%%%%%%%%%%%%%%%%%%%%%%%%%

A future task to be considered is the analysis of a more realistic
general gauge-Higgs unified model with a larger symmetry group at finite
temperature. 
The symmetry breaking in the $SU(3)$ gauge theory with various fermions
has been studied by lattice calculations
\cite{CNHH}, and it is reported that the $SU(3)$-confined phase exists at the
strong coupling regime.
We should continue to pursue the strong coupling effect and the finite size effect
with various analytical and numerical methods in order to deepen our
understanding on the Hosotani mechanism at finite temperature.   
We should also
consider general gauge theories in higher dimensional extra space, orbifold, and
warped spacetimes.

As a natural extension of the present analysis, we come to the idea that
no fermion number condition on the matter field in the model should be taken into
account. In addition, the Kaluza--Klein
charge (originated from the momentum in the extra dimension) is also a conserved
quantity \cite{KS,Shiraishi2,Shiraishi3,McGuigan,DLS}. Thus, we can also assume
the no net Kaluza--Klein charge in the closed system. 
For the case of the Hosotani mechanism, however, there appears a problematic issue
that the projection onto a state of a definite Kaluza-Klein charge breaks
the residual large gauge symmetry, such as $\theta\sim \theta+4\pi$ (or $\theta\sim
\theta+2\pi$, due to $Z_2$) in the present model. The treatment of these conserved
charges will be challenged in future with more elaborate investigations.

%%%%%%%%%%%%%%%%%%%%%%%%%%%%%%%%%%%%%%%%%%%%%%%%%%%%%%%%%%%%%%%%%
%%%%%%%%%%%%%%%%%%%%%%%%%%%%%%%%%%%%%%%%%%%%%%%%%%%%%%%%%%%%%%%%%
\appendix
%%%%%%%%%%%%%%%%%%%%%%%%%%%%%%%%%%%%%%%%%%%%%%%%%%%%%%%%%%%%%%%%%
%%%%%%%%%%%%%%%%%%%%%%%%%%%%%%%%%%%%%%%%%%%%%%%%%%%%%%%%%%%%%%%%%

\section{An approximation scheme by finite sums}

The defined functions $w_b$ and $w_f$ are expressed in the summation forms
(\ref{wb}) and (\ref{wf}).
We try to approximate these by finite sums.
We define the following function:
\begin{eqnarray}
w_{app}(\theta,\psi)&\equiv&
\frac{\Gamma(D/2)}{\pi^{D/2}}\sum_{n=1}^1\sum_{k=-2}^2\frac{\cos
k\theta
\cos n\psi} {\left[\beta^2n^2+L^2k^2\right]^{D/2}}\nonumber \\
&=&\frac{\Gamma(D/2)}{\pi^{D/2}}\cos\psi
\left[\frac{1}{\beta^D}+2\frac{\cos
\theta
} {\left(\beta^2+L^2\right)^{D/2}}
+2\frac{\cos
2\theta
} {\left(\beta^2+4L^2\right)^{D/2}}\right]\,.
\label{wapp}
\end{eqnarray}
As an approximation, we replace both $w_b$ and $w_f$ with $w_{app}$.
This approximation is justified for a large $D$ and also for $\beta/L\ll 1$.
%%%%%%%%%%%%%%%%%%%%%%%%%%%
% 05
%%%%%%%%%%%%%%%%%%%%%%%%%%%
%\begin{wrapfigure}{r}{5cm}
\begin{figure}[ht]
\centering
\includegraphics[width=5cm]
{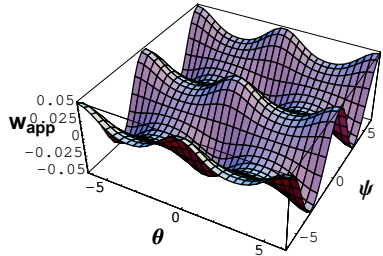}
\hspace{2cm}
\caption{The function $w_{app}$ for $\beta=1.2$ and $L=1$.}
\label{fig5}
\end{figure}
%\end{wrapfigure}
%%%%%%%%%%%%%%%%%%%%%%%%%%%
Figure \ref{fig5} shows the function $w_{app}$ for $D=5$, $\beta=1.2$, and $L=1$.
The approximation looks fine, at least for the values of parameters around this
assumption.
Owing to the simplification, we can use the following formula
to evaluate the projection integral:
\begin{equation}
\frac{1}{2\pi}\int_{-2\pi}^{2\pi}\sin^2\frac{\psi}{2}e^{
z\cos\psi}d\psi=I_0(z)-I_1(z)\,,
\end{equation}
where $I_n(z)$ is the modified Bessel function of the first kind.

According to this approximation, the sum included in the vacuum contribution in
$Z$ is also approximated by a finite sum:
\begin{equation}
\sum_{k=1}^\infty\frac{\cos k\theta}{k^D}\approx \cos\theta+\frac{\cos
2\theta}{2^D}\,.
\end{equation}

\begin{figure}[ht]
\centering
\includegraphics%[width=5cm]
{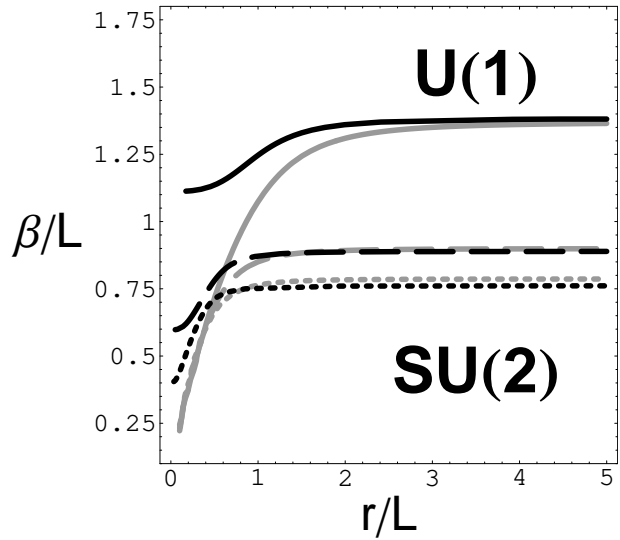}
\caption{Phase structure of the model: The lines in Fig.~\ref{fig2}
are reposted here and corresponding gray lines are drawn according to the present
approximation scheme.}
\label{fig6}
\end{figure}

Figure \ref{fig6} shows the reposted phase diagram (Fig.~\ref{fig2}) and the gray
lines obtained from the above approximation in addition.
%%%%%%%%%%%%%%%%%%%%%%%%%%%
% 06
%%%%%%%%%%%%%%%%%%%%%%%%%%%
%\begin{wrapfigure}{r}{5cm}
%\end{wrapfigure}
%%%%%%%%%%%%%%%%%%%%%%%%%%%

For a large volume, the integral over $\psi$ are dominant in the region around the
stationary points $\psi\approx 0$. Therefore the approximation of taking $n=1$
only is good for cases with large volumes.
One can see that the bending location on the lines are well approximated.

%%%%%%%%%%%%%%%%%%%%%%%%%%%%%%%%%%%%%%%%%%%%%%%%%%%%%%%%%%%%%%%%%%%%%%%%%%%
%%%%%%%%%%%%%%%%%%%%%%%%%%%%%%%%%%%%%%%%%%%%%%%%%%%%%%%%%%%%%%%%%%%%%%%%%%%
%%%%%%%%%%%%%%%%%%%%%%%%%%%%%%%%%%%%%%%%%%%%%%%%%%%%%%%%%%%%%%%%%%%%%%%%%%%

%%%%%%%%%%%%%%%%%%%%%%%%%%%%%%%%%%%%%%%%%%%%%%%%%%%%%%%%%%%%%%%%%%%%%%%%%%%
%\acknowledgments
%%%%%%%%%%%%%%%%%%%%%%%%%%%%%%%%%%%%%%%%%%%%%%%%%%%%%%%%%%%%%%%%%%%%%%%%%%%
%Acknowledgements
%%%%%%%%%%%%%%%%%%%%%%%%%%%%%%%%%%%%%%%%%%%%%%%%%%%%%%%%%%%%%%%%%%%%%%%%%%%
%\begin{acknowledgments}
%We would like to thank  for providing information on
%their work on 
%We thank
%the organizers of JGRG21, where our
%partial result %({\tt [arXiv:10mm.xxxx]}) 
%was presented. %for elucidating comments.
%This study is supported in part by the Grant-in-Aid of Nikaido Research 
%Fund.
%\end{acknowledgments}
%%%%%%%%%%%%%%%%%%%%%%%%%%%%%%%%%%%%%%%%%%%%%%%%%%%%%%%%%%%%%%%%%%%%%%%%%%%

%%%%%%%%%%%%%%%%%%%%%%%%%%%%%%%%%%%%%%%%%
%%%%%%%%%%%%%%%%%%%%%%%%%%%%%%%%%%%%%%%%%
%%%
%%%   References
%%%
%%%%%%%%%%%%%%%%%%%%%%%%%%%%%%%%%%%%%%%%%
%%%%%%%%%%%%%%%%%%%%%%%%%%%%%%%%%%%%%%%%%
%%%%%%%%%%%%%%%%%%%%%%%%%%%%%%%%%%%%%%%%%%%%%%%%%%%%%%%%%%%%%%%%%%%%%%%%%%%
%thebibliography
%%%%%%%%%%%%%%%%%%%%%%%%%%%%%%%%%%%%%%%%%%%%%%%%%%%%%%%%%%%%%%%%%%%%%%%%%%%
%\bibliographystyle{apsrev}
\bibliographystyle{apsrev4-1}
%\bibliography{}

%%%%%%%%%%%%%%%%%%%%%%%%%%%%%%%%%%%%%%%%%%%%%%%%%%%%%%%%%%%%%%%%%%%%%%%%%%%
%%%%%%%%%%%%%%%%%%%%%%%%%%%%%%%%%%%%%%%%%%%%%%%%%%%%%%%%%%%%%%%%%%%%%%%%%%%

%%%%%%%%%%%%%%%%%%%%%%%%%%%%%%%%%%%%%%%%%%%%%%%%%%%%%%%%%%%%%%%%%%%%%%%%%%%
\end{document}